\documentclass{IAU-TrB}
\usepackage{graphicx}
\newcommand{\NOPRINT}[1]{\null}

\title[PUBLIC NAMING OF PLANETS AND PLANETARY SATELLITES]     
{}

\author[EXECUTIVE COMMITTEE WORKING GROUP]   
{}

\pubyear{2015}
\volume{Volume XXIXA}
\pagerange{1--4}
\date{15 Dec 2015}
\jname{Transactions IAU, Volume XXIXA}
\editors{Thierry Montmerle, ed.}
\begin{document}

\maketitle

{\bf

\large
\begin{tabbing}
\hspace*{65mm}       \=                                              \kill
EXECUTIVE COMMITTEE \\WORKING GROUP         \> PUBLIC NAMING OF PLANETS \\ 
				\> AND PLANETARY SATELLITES  \\
\end{tabbing}

\normalsize

\begin{tabbing}
\hspace*{65mm}       \=                                              \kill
CHAIR            \> Thierry Montmerle$^\star$      \\
MEMBERS			 \> Piero Benvenuti,$^\star$ Sze-leung Cheung,    \\
					\>  Lars Lindberg Christensen,  \\
					\>  Alain Lecavelier des Etangs,  \\
					\>  Xiaowei Liu,$^\star$ Donald Lubowich  \\
					\>  Eric Mamajek, Rita Schulz  \\
					\>  Giovanni Valsecchi, Gareth Williams   \\
					\>  Robert Williams$^\star$  \\

\end{tabbing}
}

\small
\noindent
\hspace*{65mm} \rule{30mm}{0.4pt}  \\
\hspace*{65mm} $^\star$ Executive Committee Member  \\

\normalsize

\bigskip
\noindent
{\bf\large FIRST REPORT (2013-2015)} 



\section{Context and creation of the Working Group}

While one of the IAU's missions is to ``serve as the internationally recognized authority for assigning designations to celestial bodies and surface features on them" (\footnote {See ``About the IAU" in the IAU web site, {\tt http://www.iau.org/about/}, and Appendix 1}), the participation of the public in the naming of celestial objects has been a little-known, but decade-long tradition of the IAU. 

Until recently this participation had several restrictions:

\begin{itemize}

\item So far, only Solar System objects were concerned (minor planets, planetary satellites, planetary surface features according to their topographical types) (\footnote {See {\tt http://planetarynames.wr.usgs.gov/DescriptorTerms}});

\item The names suggested by the public were put in a name bank, and extracted in due time by IAU specialized task groups according to certain rules (example: Greek or Roman mythological names):

\hspace*{7mm}-- Division F {\it Working Group on Small Bodies Nomenclature} (WGSBN: {\tt http://www.\\iau.org/science/scientific\_bodies/working\_groups/97/}), supporting the Minor \\ Planet Center (MPC: {\tt http://www.minorplanetcenter.net/});

\hspace*{7mm}-- Division F {\it Working Group on Planetary System Nomenclature} (WGPSN: {\tt http://\\www.iau.org/science/scientific\_bodies/working\_groups/98/})  \\
	
\end{itemize}

To date, names have been attributed by the IAU to $\sim 20\,000$ minor planets (see {\tt http://www.minorplanetcenter.net/iau/lists/MPNames.html}), and to over $9\,000$ lunar craters and features ({\tt http://planetarynames.wr.usgs.gov/SearchResults?tar\\get=MOON}). Among those, a few hundred have been drawn from public suggestions. \\

Nevertheless, this IAU ``monopoly" on the naming of celestial objects has been challenged by some groups, which invited the public to send them directly names of celestial objects, almost without screening or restriction. The most vocal group has been the {\it Uwingu} organization, which started in 2013 soliciting names from the public for Mars craters and exoplanets for a fee, as a crowd-sourcing means ``to fund science education" ({\footnote {\tt http://www.uwingu.com/}}).

The IAU expressed its position in a Press Release against ``Buying the Right to Name a Planet'' $[1]$. 
While the IAU was on firm ground because of its mission in the naming of Mars craters (which continues as science requests it)(\footnote {See Appendix 1; the current $> 1,000$ names of Mars craters can be found at  {\tt http://plane\\tarynames.wr.usgs.gov/SearchResults?target=MARS\&featureType=Crater,\%20craters}}), the question of exoplanet naming was debated regularly within the IAU Commission 53, but no consensus was reached on whether ``names" should be added to the exoplanet designation. However, in 2012 the Commission favored the designation first introduced by the Geneva group, inspired from the tradition of multiple star system nomenclature (i.e., appending upper-case letters: ``A" for the primary, ``B", ``C..." for the companions) by appending the lower-case letters ``b, c..." (in order of discovery) to the IAU-compliant designation of the host star. 

However, in view of the rapid pace of exoplanet findings (about 800 were already confirmed
 in 2013), and of their potential links with the search for life in the universe,  the public complained, via the press or social networks, that exoplanets they heard about were designated only by ``telephone" or ``license-plate" numbers (think of JMASS J2126-8140, or OGLE-2008-BLG-355L b...), contrary to Solar System objects for example. In a way, the {\it Uwingu} initiative built on these complaints and openly criticized the IAU; 20 years after the first discoveries, the public did not understand why no name was given to exoplanets, and, more importantly, why it could not be associated with some naming process.


In response to this situation, and while reiterating its opposition to having the public pay to give a name to an exoplanet, the IAU Executive Committee nonetheless recognized the right of organizations to invite public, international exoplanet naming and/or voting campaigns. To this end, clear selection rules were to be defined by the IAU, inviting mutual collaboration, the goal being to sanction the campaign and officially approve the resulting names $[2][3]$, for the sake of boosting the public's interest in astronomy and at the same time reaffirm the authority of the IAU. In no way were these names supposed to supersede the designations in use by professional astronomers.

In order to review public naming campaigns by organizations, not only concerning exoplanets but also Solar System objects, the Executive Committee, at its EC93 meeting in Nara (2013) and at the initiative of the IAU General Secretary, decided as a first step to create a small ``Executive Committee Task Group'' (EC-TG) on {\it Public Naming of Planets and Planetary Satellites}.  The first set of naming rules published $[2]$ was meant to apply to exoplanets but were very similar to those already in force for Solar System objects, and emphasized the condition of a free, worldwide process.

Its composition was as follows:

\begin{itemize}

\item {\bf Science expert:}

\hspace*{7mm}-- Division F (Planetary Systems and Bioastronomy): G. Valsecchi (Div.F President)



\item {\bf Media:}

\hspace*{7mm}-- IAU Press Office: L. L. Christensen (Press Officer)

\item {\bf Executive Committee members:}


\hspace*{7mm} Xiaowei Liu (Vice-President)

\hspace*{7mm} T. Montmerle (GS, Chair)

\hspace*{7mm} R. Williams (Past President)

\end{itemize}

The Press Office was included from the start, as it was anticipated that any action by the IAU concerning or approving the public naming of Solar System objects and exoplanets would have a high visibility worldwide, and trigger many Press Releases and Announcements ---as well as reactions from the press and the public.


\section{Status of public naming of Solar System objects}

As mentioned above, the IAU has a long tradition, not only of using name suggestions by individuals, but also of cooperating with public institutions like NASA. 

First, the rules for naming astronomical objects by individuals, in particular Solar System objects like asteroids and minor planets, have been defined for a long time by the MPC, and can be found in a ``theme" accessible from the IAU home page ({\tt http://www.\\iau.org/public/themes/naming/}).

Second, the IAU also has a long tradition of collaboration with space agencies for naming celestial objects discovered by space missions.  However, the public was generally not involved in the naming process until recently. A notable exception was the organization by NASA, long before the Internet era, of a public competition to name the cloud-covered surface features of Venus discovered in the course of the {\sl Magellan} mission (1989--1994): in agreement with the IAU, the proposed theme was ``famous women", both mythical and real (the only exception being James Clerk Maxwell, nominated in reference to the radar carried by the probe)(\footnote {\tt http://www2.jpl.nasa.gov/magellan/guide8.html}).

More recently, the IAU was involved in or sanctioned various naming campaigns for planetary satellites and surface features (typically name suggestions by the public followed by public votes to select the most popular names): 

\begin{itemize}

\item The two latest Pluto satellites (originally known as P3 and P4) discovered by {\sl HST} [4] (2013);

\item The last craters discovered on Mercury at the end of the {\sl Messenger} survey mission [5][6] (2014--2015);

\item Pluto surface features, discovered by the {\sl New Horizons} spacecraft [7] (2015; still pending). 

\end{itemize}

However, in all the examples mentioned above, while the IAU was setting the rules, other organizations (here NASA) actually took the initiative of directly involving the public in the naming campaigns, and organized the votes themselves. (For the record, ESA also approached the IAU in 2014 for the public naming of surface features on comet 67P/Churyumov-Gerasimenko in the course of the {\sl Rosetta/Philae} mission, but the IAU declined because these features are not permanent.) 

Since the field of exoplanet research was still ``virgin" for naming but becoming scientifically mature (with hundreds of well-characterized, confirmed objects) and so appealing to the public's interest and imagination, any appropriate organization, private or public, could have started a comparable naming campaign for exoplanets. But on the other hand, with its in-house experience  and scientific structure (Div.F, C53, WGPSN, WGSBN/MPC, etc.), the IAU could simply decide to organize such a naming contest itself with the best possible cards in its hands, and be the leader of a new project with a potentially high worldwide impact, the first since the International Year of Astronomy in 2009.

This is how the {\it NameExoWorlds} project came to life.


\section{Organizing the public naming of exoplanets}

Strongly inspired by the participation of the IAU in campaigns for naming Solar System objects, the EC-TG started to draw up a project for the public naming of exoplanets.

Several specific features were introduced:

\begin{itemize}

\item {\it The naming process would be restricted to astronomy clubs and non-profit associations interested in astronomy} (hereafter ``clubs'' for short). The definition of ``associations" was to be understood in a broad way, and could include schools, historical or cultural associations, etc.  Naming by individuals was not to be allowed: a two-week long pilot project opening of the IAU Secretariat {\tt iaupublic@iap.fr} email address to exoplanet naming suggestions in September 2013 collected nearly 2\,000 responses, but of low quality (self-nominations, pet names, lists of names obviously copied from dictionaries, etc.). It was thought that restricting the participation to clubs would spur discussions within the interested groups, be more international, and have a strong educational value.

\item The clubs would have to first register in a dedicated, publicly accessible {\it IAU Directory for World Astronomy}. This Directory would be set up by the newly created {\it Office of Astronomy Outreach} (OAO) in Tokyo. 

This approach would have two main advantages:

\begin{itemize}

\item One could check the good standing of the clubs (in contrast to individuals, who could have more than one email address), monitor their membership, and assess their geographical distribution;

\item The Directory would become the backbone of a permanent, worldwide IAU-led network that could be involved in future projects of the OAO and other IAU structures (such as the Office of Astronomy for Development: OAD, in Cape Town).

\end{itemize}

\item The IAU itself would provide a sample of {\it bona fide}, well-characterized exoplanetary systems, with the agreement of the discoverers and the exoplanet community (represented by Division F Commission 53). This sample would satisfy the 2003 ``Working Definition of a Planet" elaborated by Division III {\it Working Group on
Extrasolar Planets} (which became Commission 53 in 2006) :
``Objects with true masses below the limiting mass for thermonuclear
fusion of deuterium (currently calculated to be 13 Jupiter masses for
objects of solar metallicity) that orbit stars or stellar remnants are
`planets' (no matter how they formed). The minimum mass/size required
for an extrasolar object to be considered a planet should be the same
as that used in our Solar System." 

\item Their host stars would be also offered for public naming, unless they already had well-known, historic popular names (Arabic, Greek, Roman, etc.);

\item This sample would be ranked by a vote organized by the OAO among the registered clubs, and the top $N$ systems offered for naming proposals by these clubs ($N$ being decided {\it a posteriori} by the TG as a function of the size of the exoplanet sample and the number of clubs having registered);

\item Names would be proposed for an entire system at once (somehow in analogy with the Solar System), e.g., 4 names for a system of 1 ``nameable" star and 3 exoplanets;

\item Each club could propose only one set of names (host star and planets), for only one system;

\item The worldwide public would be solicited to vote on the club proposals sent to the IAU;

\item The names included in the top-voted proposals (``the winning names") would be officially sanctioned by the IAU, and recorded in professional databases like SIMBAD.  \\

\end{itemize}

Implementing such a project required new expertise to be added to the EC-TG. 

First, a new Working Group was established, under Div.F, entitled {\it Exoplanets for the Public} ({\tt http://www.iau.org/science/scientific\_bodies/working\_groups/211/}), and chaired by A. Lecavelier des Etangs (C53 President). This Working Group undertook the task of contacting the community and defining the sample of exoplanets available for public naming. 

Then, after preparatory telecons and exchange of working documents over several months, a founding meeting, hosted by G. Valsecchi and chaired by T. Montmerle, took place on the INAF premises of Tor Vergata (Rome, Italy), on 9--10 April 2014, just before the EC94 Executive Committee Meeting in Canberra (Australia). In addition to the EC-TG members, this meeting included representatives of the newly created Div.F-WG (A. Lecavelier des Etangs), the WGPSN (R. Schulz), the WGSBN/MPC (G. Williams), the IAU Press Office (L. L. Christensen), the OAO (Sze-leung Cheung), and also P. Benvenuti (AGS) as an additional EC member. While R. Williams (in Baltimore) and Xiaowei Liu (in Beijing) could be available only separately, C. Lintott (from the Zooniverse organization) joined the meeting by telecon, with the perspective of a joint ``citizen science" approach to the exoplanet naming project, still without a name.   \\

The main decisions coming out of this meeting were as follows:

\begin{itemize}

\item From the WG {\it Exoplanets for the Public}, a sample of 305 {\it bona fide}, well-characterized exoplanets, discovered before 31 December 2008 (representing 260 systems of one to five exoplanets, studied for more than five years), was made available for public naming;

\item The Task Group would be ``upgraded" to an EC-Working Group by including the meeting participants listed above, and proposed for approval to the EC in Canberra (to this list would be added in 2015 a representative of the Division B Commission 5 WG on {\it Designations}, for the star names: D. Lubowich, as well as another member of the new WG: E. Mamajek). (More details on the composition of this EC-WG are given in Appendix 2); 

\item The project, as outlined above (and adding budget details), was agreed by the participants, and submitted to the EC for approval;

\item A MoU was to be drafted by T. Montmerle (as GS) and agreed between the IAU and Zooniverse: the {\it Directory for World Astronomy} was to be developed by the OAO, while the adaptation of their registration and voting software to IAU/OAO requirements, as well as the final, worldwide voting, was to be implemented and organized by Zooniverse.  \\

\end{itemize}


\section{NameExoWorlds: from project to contest}

The creation of the EC-WG (with the same name as the pathfinder EC-TG: {\it Public Naming of Planets and Planetary Satellites}) was approved at the EC94 meeting in Canberra (30 April--2 May 2014), with the mission to organize the two projects leading to the first ``public naming of exoplanets and their host stars": 

$(i)$ the creation of the IAU {\it  Directory for World Astronomy} ({\tt https://directory.iau.\\org/}), designed by the OAO (which was to win the 2015 Best Science Website Award [8]), and 

$(ii)$ the contest, now called {\it NameExoWorlds}, an ``ExoWorld" being defined as an exoplanetary system {\it and} its host star.  \\

The official ``birth" of the contest was 9 July 2014, when the IAU issued a Press Release [9] announcing the opening of its web site, designed by Zooniverse under the supervision of the OAO ({\tt http://nameexoworlds.iau.org/}), that included the sample of 305 exoplanets along with related properties and explanations, the steps and timeline of the process, the rules and guidelines for proposing names and for voting eligibility. The timeline was originally set so as to announce the ``winning names" at a special ceremony being held during the Honolulu GA a year after.   \\

The following steps were accomplished, in conformity with this timeline: 

\begin{itemize}

\item {\it Opening of registration} in the Directory for World Astronomy: 3 October 2014 $[10]$. The registration was originally planned to last three months, but was eventually kept open indefinitely, so that the membership could grow for future projects (the Directory now has nearly 700 member clubs);

\item {\it Vote by the registered clubs} to rank the most interesting systems: 13 January 2015 $[11]$. The Directory comprising $\sim 400 $ clubs at the time, the EC-WG chose to select the top 20 systems for naming, and fed them into the {\it NameExoWorlds} web page. The resulting list was surprisingly diverse in scientific interest, including in particular the first exoplanetary system, discovered in 1990 around the pulsar PSR 1257+12, the five-exoplanet system 55 Cnc, the exoplanet/circumstellar disk system Fomalhaut, etc. Obviously the clubs had been very careful in their choices, and the outcome demonstrated that engaging clubs, rather than individuals, into naming exoplanets was a very fruitful approach. 

\item {\it Selection of the top 20 ExoWorlds}, representing 32 exoplanets and 15 ``nameable" stars, i.e., not already having a popular name. Here, the WG considered that the names for the host stars $\epsilon$ Tau (Ain), $\iota$ Dra (Edasich), $\gamma$ Cep (Errai), $\alpha$ PsA (Fomalhaut), and $\beta$ Gem (Pollux) were in sufficiently common usage in the astronomical and historical literature to be adopted as official, so that only their exoplanets would be open for naming via NameExoWorlds.

\item {\it Opening of submission of naming proposals} by the registered clubs: 27 April 2015 $[12]$. A period of about three months was left to the clubs so select their favorite ExoWorld (as mentioned above only one proposal was allowed per club), and the deadline for closing the submission was set to 15 June. A total of 237 proposals were eventually received from 45 countries, for a total of 551 distinct names.  \\

\end{itemize}

Then the contest entered  a crisis. Communication problems with Zooniverse, the sudden departure of their web developer in early June, etc., brought the project to a halt. The MoU was never signed, and it was discovered that the proposals received had not been properly stored in the Zooniverse system ! With the Honolulu GA only a few weeks away, the contest looked doomed. But swift action by Sze-leung Cheung and the OAO followed: a new contract was established with the software developers who had designed the Directory, and within a couple of weeks only the proposals were safely recovered from the proposers themselves, the contest put back on track and the worldwide voting process organized. However, it was clear that the vote could not be completed in time for the GA ---so it was decided to {\it open} it at the GA, instead of announcing the results.

Eventually the voting was successfully opened on 12 August during a special public session of the GA, kindly chaired by Lisa Kaltenegger (Director of the Carl Sagan Institute at Cornell) $[13]$, with a closing deadline of 31 October, not far from the 20th anniversary of the publication of the discovery of the first exoplanet orbiting a solar-type star by M. Mayor and D. Queloz (51 Peg b: Nature 378, 355; 23 Nov. 1995; of course part of the 20 ExoWorlds to be named).


\section{Results and lessons to be learned}

The results (the ``winning names" and the ``winning clubs") were announced on 15 December 2015 $[14]$, i.e., six weeks after the counting of the votes had become available. 

$(i)$ The first reason for such a relatively long period was the sheer amplitude of the worldwide vote. The raw number of votes was 631\,418. An analysis of the votes received as a function of time however revealed anomalies that could be attributed to multiple vote fraud (only one vote per ExoWorld was allowed). The OAO made a careful analysis, which led to removing $\sim 10\%$ of the votes (58\,176). In the end, the number of valid votes has been 573\,242, coming from 182 countries and territories (see {\tt http://nameexoworlds.iau.org/statistics}).

$(ii)$ The biggest stumbling block has been the fact that some of the winning names were already attributed to other celestial objects, in most cases asteroids or minor planets, which was undesirable. These duplicates had not been weeded out before the vote. Extended discussions took place within the EC-WG, sometimes including outside experts, to decide which names to adopt. In the majority of cases, and after negotiations on a case-by-case basis with the proposers respecting the spirit of their proposals, the duplicate names could be adapted before being officially accepted. This phase was most challenging, but very rich in exchanges about history and cultures.

$(iii)$ Another difficulty arose from the fact that one of the winning names (for the $\tau$ Bo\"{o} system) turned out to be that of a person not satisfying the IAU requirement of not being ``principally known for political, military or religious activities". The vote was consequently removed from the contest, so in the end only 19 ExoWorlds were named.  \\

Checking for astronomical duplicates and controversial names before launching the vote will obviously be the first prerequisite if another {\it NameExoWorld} contest is to be organized in the future.  

In spite of these hiccups, the final list of approved names $[14]$ is quite remarkable (see Appendix 3). It is composed of a wide variety of names taken from mythology, literature, or are even pure creations, but somewhat unexpectedly there were no proposal from science-fiction lore. In line with the IAU culture, the approved names represent all the continents, and there is no strong bias towards a particular country or world region. Four winning proposals were received from North America (USA, Canada), one from Latin America and the Caribbean (Mexico), two from the Middle East \& Africa (Morocco, Syria), six from Europe (France, Italy, Netherlands, Spain, Switzerland), and six from Asia--Pacific (Australia, Japan, Thailand). Also, winning proposals came not only from astronomy clubs, but from schools and universities as well.

In addition, from private sources we have significant evidence that the contest triggered local actions, or even lobbying, by clubs to promote astronomy towards a broader public.

A last -- but not least -- consideration was the press coverage during the contest: an OAO web survey showed that over 800 articles were published worldwide, from 54 countries, spread about equally between the pre-results and the post-results periods. 


Altogether, we would conclude at this stage that the {\it NameExoWorlds} project has undeniably had a great public impact, in line with the long-standing successful tradition of IAU global initiatives in outreach and education.  \\

\newpage


{\bf References: IAU Press Releases and Announcements relevant to the NameExoWorlds project}


\begin{table}[h]
\begin{center}
\begin{tabular}{ccll}

\hline
{\it Ref. }&{\it Date}  & {\it Title} &{\it URL}$^{(1)}$  \\
\hline

$[1]$	&	12 Apr 2013	& 	{\sf Can One Buy the Right to Name a Planet ?}	& PR {\tt /iau1301/} 	\\
	&		&	{\sf -- The IAU Responds to Recent Name-Selling Campaign} &   \\
$[2]$	&	14 Aug 2013	&	{\sf Public Naming of Planets and Planetary Satellites}  &  Ann {\tt /ann13009/}  \\
$[3]$	&	18 Dec 2013	&	{\sf The Naming of Exoplanets}	&	Ann {\tt /ann13012/}  \\
$[4]$	&	02 Jul 2013	&	{\sf Names for New Pluto Moons Accepted by the IAU After}  	& PR  {\tt /iau1303/}  \\
	&				&	{\sf Public Vote} &  \\
$[5]$	&	18 Dec 2014	&     {\sf Public Contest to Name Craters on Planet Mercury} & PR {\tt /iau1407/} \\
$[6]$	&	29 Apr 2015	&	{\sf Mercury Crater-naming Contest Winners Announced} & PR {\tt /iau1506/} \\
$[7]$	&	24 Mar 2015	&	{\sf Campaign for Public Participation in Naming Features} &	PR {\tt /iau1502/}  \\
	&				&	{\sf on Pluto}  &  \\
$[8]$	&	04 Sep 2015	&	{\sf IAU Directory of World Astronomy wins 2015 Best}  	&	Ann {\tt /ann15029/}  \\
	&				&	{\sf Science Website Award}	&  \\
$[9]$	&	09 Jul 2014	&	{\sf NameExoWorlds: An IAU Worldwide Contest to Name}   & PR {\tt /iau1404/}  \\
	&				&	{\sf Exoplanets and their Host Stars}  &  \\
$[10]$	&	03 Oct 2014	&	{\sf Register Now to Enter NameExoWorlds Contest}  &  PR {\tt /iau1406/}  \\
	&				&	{\sf -- IAU invites astronomical organisations to join the } &  \\
	&				&	{\sf Directory for World Astronomy } &  \\
$[11]$ &	13 Jan 2015	&	{\sf NameExoWorlds Contest Opens -- Propose your}  &  PR {\tt /iau1501/}  \\
	&				&	{\sf favourite exoplanetary system now}  &  \\
$[12]$ &	27 Apr 2015	&	{\sf 20 ExoWorlds are now available for naming proposals}  &  PR {\tt /iau1505/}  \\
$[13]$  &	12 Aug 2015	&	{\sf NameExoWorlds Contest Opens for Public Voting}  &  PR  {\tt /iau1511/}  \\
	&				&	{\sf -- Vote now for your favourite names from the IAU's shortlist}  &  \\
$[14]$ &	15 Dec 2015	&	{\sf Final Results of NameExoWorlds Public Vote Released}  &  PR  {\tt /iau1514/}  \\

\hline

\end{tabular}
\end{center}
\end{table}

\noindent
(1) Format: ``PR" = Press Release; ``Ann" = Announcement; ``/xxxNNNN/" is short for   \\ {\tt http://www.iau.org/news/pressreleases/detail/iauNNNN/}, and \\ {\tt http://www.iau.org/news/announcements/detail/annNNNNN/}, respectively

\newpage

\bigskip
\bigskip

{\bf Appendix 1: IAU role in naming planetary objects}  \\

Through the recognition of national committees, academies, and
  governments, the IAU has been the international authority on celestial
  nomenclature since its inception in 1919. Among the IAU's early
  activities were the international standardization of the nomenclature
  of surface features on the Moon (1935) and Mars (1958). Since 1973,
  planetary nomenclature has been handled by the IAU Division Working Group for
  Planetary System Nomenclature (WGPSN; formerly under Div.III, now under Div. F), with individual task groups
  assigned to planets and classes of objects (e.g. Moon, Mercury, Mars,
  Outer Solar System, Small Bodies).

\bigskip
\bigskip

{\bf Appendix 2: Composition of the EC-WG ``Public Naming of Planets and Planetary Satellites" (end 2015):}   \\

To capitalize on the experience of the IAU in the public naming of Solar System objects (as explained in the text), its current composition, showing in particular a strong involvement of Div.F, is as follows:  \\

\begin{itemize}

\item {\bf Science experts:}

\hspace*{7mm}-- Division F {\it Planetary Systems and Bioastronomy}: G. Valsecchi (past Div.F President)

\hspace*{7mm}-- Div.F/WGPSN: R. Schulz (Chair WG)

\hspace*{7mm}-- Div.F/WGSBN/MPC: G. Williams (MPC Associate Director)

\hspace*{7mm}-- Div.F/WG {\it Exoplanets for the Public}: A. Lecavelier des Etangs (Chair WG; Commission F2 President); E. Mamajek

\hspace*{7mm}-- Div.B/C.B2/WG {\it Designations} (stars): D. Lubovich  \\

\item {\bf Media/Outreach:}

\hspace*{7mm}-- IAU Press Office: L. L. Christensen (Press Officer)

\hspace*{7mm}-- Office of Astronomy Outreach (OAO): Sze-leung Cheung (International Outreach Coordinator)  \\

\item {\bf Executive Committee members:}

\hspace*{7mm}-- P. Benvenuti (GS)

\hspace*{7mm}--  Xiaowei Liu (Vice-President)

\hspace*{7mm}--  T. Montmerle (past GS, EC-WG Chair)

\hspace*{7mm}--  R. Williams (Past President)  \\

\end{itemize}

\bigskip
\bigskip

{\bf Appendix 3: Stellar and planetary names officially adopted as a result of the NameExoWorlds contest }  \\

(Citations and club names can be found at {\tt http://nameexoworlds.iau.org/names}).

\begin{table}[h]
\begin{center}
\begin{tabular}{lllcl}

\hline
{\it Star/Planet }&{\it Designation}  & {\it Adopted name} &{\it Country} & {\it Club/Association} \\
\hline

Star		&	14 And		& 	Veritate		&	Canada	&	Astronomy Club	\\
Planet 	&	14 And b	 	&	Spe			&			&					\\
 	&		 	&				&			&					\\

Star		&	18 Del		& 	Musica		&	Japan	&	High School	\\
Planet 	&	18 Del b 		&	Arion			&			&					\\
	&		 	&				&			&	 				\\

Star		&	42 Dra		& 	Fafnir		&	USA		&	Astronomy Club	\\
Planet 	&	42 Dra b 		&	Orbitar		&			&					\\
	&		 	&				&			&	 				\\
	
Star		&	47 UMa		& 	Chalawan		&	Thailand		&	Astronomy Club	\\
Planet 	&	47 UMa b 		&	Taphao Thong	&			&					\\
Planet 	&	47 UMa c 		&	Taphao Kaew	&			&					\\
	&		 	&				&			&	 				\\
	
Star		&	51 Peg		& 	Helvetios		&	Switzerland	&	Astronomy Club	\\
Planet 	&	51 Peg b 		&	Dimidium		&			&					\\
	&		 	&				&			&	 				\\
	
Star		&	55 Cnc		& 	Copernicus	&	Netherlands	&	Astronomy Club	\\
Planet 	&	55 Cnc b 		&	Galileo	&			&					\\
Planet 	&	55 Cnc c 		&	Brahe	&			&					\\
Planet 	&	55 Cnc d 		&	Lipperhey	&			&					\\
Planet 	&	55 Cnc e 		&	Janssen	&			&					\\
Planet 	&	55 Cnc f 		&	Harriot	&			&					\\
	&		 	&				&			&	 				\\

Planet 	&	Ain b	 ($\epsilon$ Tau b)	&	Amateru	&	Japan	&	Astronomy Club	\\
 	&		 	&				&			&					\\

Planet 	&	Edasich b	 ($\iota$ Dra b)	&	Hypatia	&	Spain	&	Student Association	\\
 	&		 	&				&			&					\\

Star		&	$\epsilon$ Eri		& 	Ran		&	USA		&	Middle School	\\
Planet 	&	$\epsilon$ Eri b 	&	AEgir	&			&					\\
	&		 	&				&			&	 				\\

Planet 	&	Errai b ($\gamma$ Cep b)	&	Tadmor	&	Syria 	&	Astronomy Club	\\
 	&		 	&				&			&					\\

Planet 	&	Fomalhaut b ($\alpha$ PsA b)	&	Dagon	&  USA 	&	Astronomy Club	\\
 	&		 	&				&			&					\\

Star		&	HD 104985		& 	Tonatiuh		&	Mexico	&	Astronomy Club	\\
Planet 	&	HD 104985 b 		&	Meztli		&			&					\\
	&		 	&				&			&	 				\\

Star		&	HD 149026		& 	Ogma		&	France	&	Astronomy Club	\\
Planet 	&	HD 149026 b 		&	Smertrios		&			&					\\
	&		 	&				&			&	 				\\

Star		&	HD 81688			& 	Intercrus		&	Japan	&	Astronomy Club	\\
Planet 	&	HD 81688 b 		&	Arkas		&			&					\\
	&		 	&				&			&	 				\\

Star		&	$\mu$ Ara			& 	Cervantes		&	Spain	&	Astronomy Club	\\
Planet 	&	$\mu$ Ara b 		&	Quijote	&			&					\\
Planet 	&	$\mu$ Ara c 		&	Dulcinea	&			&					\\
Planet 	&	$\mu$ Ara d 		&	Rocinante	&			&					\\
Planet 	&	$\mu$ Ara e 		&	Sancho	&			&					\\
	&		 	&				&			&	 				\\

Planet 	&	Pollux b ($\beta$ Gem b)	&	Thestias	&  Australia 	&	Astronomy Club	\\
 	&		 	&				&			&					\\

Star		&	PSR 1257+12		& 	Lich		&	Italy	&	Astronomy Club	\\
Planet 	&	PSR 1257+12 b 	&	Draugr	&			&					\\
Planet 	&	PSR 1257+12 c 	&	Poltergeist	&			&					\\
Planet 	&	PSR 1257+12 d 	&	Phobetor	&			&					\\
	&		 	&				&			&	 				\\

Star		&	$\upsilon$ And		& 	Titawin		&	Morocco	&	Astronomy Club	\\
Planet 	&	$\upsilon$ And b 	&	Saffar		&			&					\\
Planet 	&	$\upsilon$ And c 	&	Samh		&			&					\\
Planet 	&	$\upsilon$ And d 	&	Majriti		&			&					\\
	&		 	&				&			&	 				\\

Star		&	$\xi$ Aql			& 	Libertas		&	Japan	&	Student Association	\\
Planet 	&	$\xi$ Aql b 		&	Fortitudo		&			&					\\
	&		 	&				&			&	 				\\

\hline

\end{tabular}
\end{center}
\end{table}

\end{document}